# Accuracy of the analytical escape rate for a cusp barrier in the overdamping regime

A V Zakharov[1], M V Chushnyakova[2], I I Gontchar[3]

[1] Pre-university Training Department, Omsk State Technical University, Omsk 644050, Russia
[2] Physics Department, Omsk State Technical University, Omsk 644050, Russia
[3] Physics and Chemistry Department, Omsk State Transport University, Omsk 644046, Russia

maria.chushnyakova@gmail.com

**Abstract**. For the first time, the accuracy of the approximate analytical Kramers formula for the thermal decay rate over a cusp barrier, $R_K$, is checked numerically for the overdamping regime. The numerical quasistationary rate, $R_D$, which is believed to be exact within the statistical errors is evaluated by means of computer modeling of the stochastic Langevin-type dynamical equations. The agreement between $R_K$ and $R_D$ significantly depends upon the friction strength and the height of the barrier in comparison to the thermal energy. The difference between $R_K$ and $R_D$ decreases with the dimensionless damping parameter $\varphi$, however, does not become smaller than 10-20%. The unexpected growth of the difference between $R_K$ and $R_D$ with the governing parameter is observed.

## 1. Introduction

The problem of the thermal decay of a metastable state (escape of a Brownian particle from a trap due to thermal fluctuations) is significant for different branches of natural sciences: in chemistry [1–3], biophysics [4,5], astrophysics [6], electronics [7,8], nuclear physics [9,10] etc. The decay rate is the principal characteristics of this process. Approximate analytical formulas for the rate accounting for dissipation were derived by Kramers in his seminal work [11]. Thus, the problem is often referred to as the Kramers problem [12,13]. In [14] it was shown that the decay rate depends mostly upon two dimensionless parameters, namely a governing parameter $G$ and damping parameter $\varphi$. The definitions of these parameters read

$$G = \frac{U_b}{\theta}, \qquad (1)$$

$$\varphi = \frac{\eta \tau_c}{2\pi m}. \qquad (2)$$

Here $U_b$ is the height of the potential barrier separating the trap from the domain of infinite motion; $\theta$ is the average thermal energy corresponding to one vibrational degree of freedom; $\tau_c$ is the oscillation period of the Brownian particle near the bottom of the potential well; $m$ and $\eta$ are the inertia and friction parameters. Depending on the value of the damping parameter they mark the energy diffusion regime ($\varphi \ll 1$), the overdamping regime (the spatial diffusion regime, $\varphi \gg 1$) and the intermediate regime (the phase space diffusion regime, $\varphi \cong 1$).

Accuracy of the Kramers formulas for the decay rate derived in [11] was studied in several articles (see, e.g., [15–17]). Unfortunately, sometimes the Kramers formulas were misinterpreted (see, for example, [18–20] and detailed discussion of this problem in [21,22]). Usually, they consider the Kramers formulas obtained for the parabolic barrier. Meanwhile, there is an approximate formula obtained by Kramers for a cusp-shaped barrier (see Fig. 2 and an equation after this figure in [11]). By our knowledge, the accuracy of this formula was not studied so far. This is the goal of the present work to fill this lacuna.

## 2. Model

In [11] Kramers considered the one-dimensional motion of the Brownian particle, therefore we also perform the modeling for a single dimensionless coordinate $q$ and its conjugate momentum $p$. The numerical scheme for modeling the Langevin equations read:

$$p^{(n+1)} = p^{(n)}(1 - \eta m^{-1}\tau) + K^{(n)}\tau + g b^{(n)}\sqrt{\tau}, \tag{3}$$

$$q^{(n+1)} = q^{(n)} + (p^{(n)} + p^{(n+1)})\tau/(2m). \tag{4}$$

Here the superscripts $n$ and $n + 1$ correspond to two successive time moments separated by the time step of the modeling $\tau$. In (3), $K = -dU/dq$ is the conservative force, the last term represents the random force; $b^{(n)}$ denotes the random number possessing the Gaussian distribution with zero average and variance 2. The value $g = \sqrt{\theta \eta}$ is the amplitude of the random force.

The potential profile used in the present work is shown in Fig. 1. The potential $U(q)$ is composed of two parabolas of the same stiffness matched at $q_b = 1.6$:

$$U(q) = \begin{cases} C(q - q_c)^2/2 & \text{at } q < q_b; \\ C(q - (2q_b - q_c))^2/2 & \text{at } q > q_b. \end{cases} \tag{5}$$

Here the subscript "$c$" refers to the bottom of the left well, subscript "$b$" indicates the barrier top, $C$ denotes the stiffness, defined by the values of $q_c$, $q_b$ and by the height of the barrier, $U_b$ (see Fig. 1). The trajectories start at $q_c = 1.0$ with zero momentum. The value of $q_a = 2.2$ (the bottom of the right well) corresponds to the absorbing border.

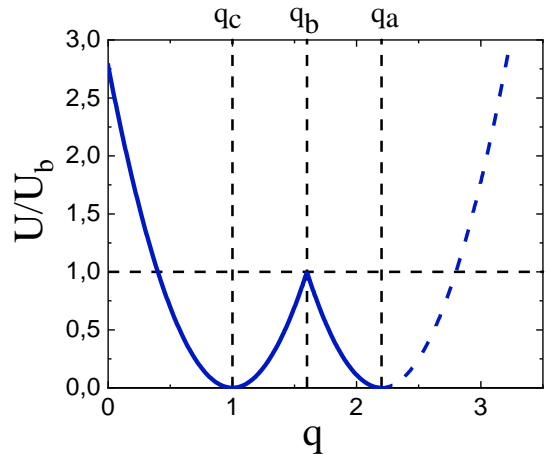

**Figure 1.** The dimensionless cusp potential.

We model $N_{tot}$ trajectories, each no longer than $t_D$. Some trajectories reach the absorbing border at $t_a < t_D$. These trajectories correspond to the decay of the metastable state. Two typical trajectories (i.e. the time dependence of the coordinate, momentum, and total energy) are displayed in Fig. 2. Since the random term enters the differential equation for the momentum (3), the time evolution of the momentum of the Brownian particle (Fig. 2a,d) is saw-edged whereas the evolution of the coordinate is much more smooth. The left column of Fig. 2 represents the trajectory corresponding to the decay of the metastable state: at the time moment $t \approx 2.1\tau_c$, the coordinate achieves the absorbing border. Before this, the particle crosses the bottom of the left well several times. Its total energy achieves the value equal to the potential barrier height several times before the particle overcomes the barrier.

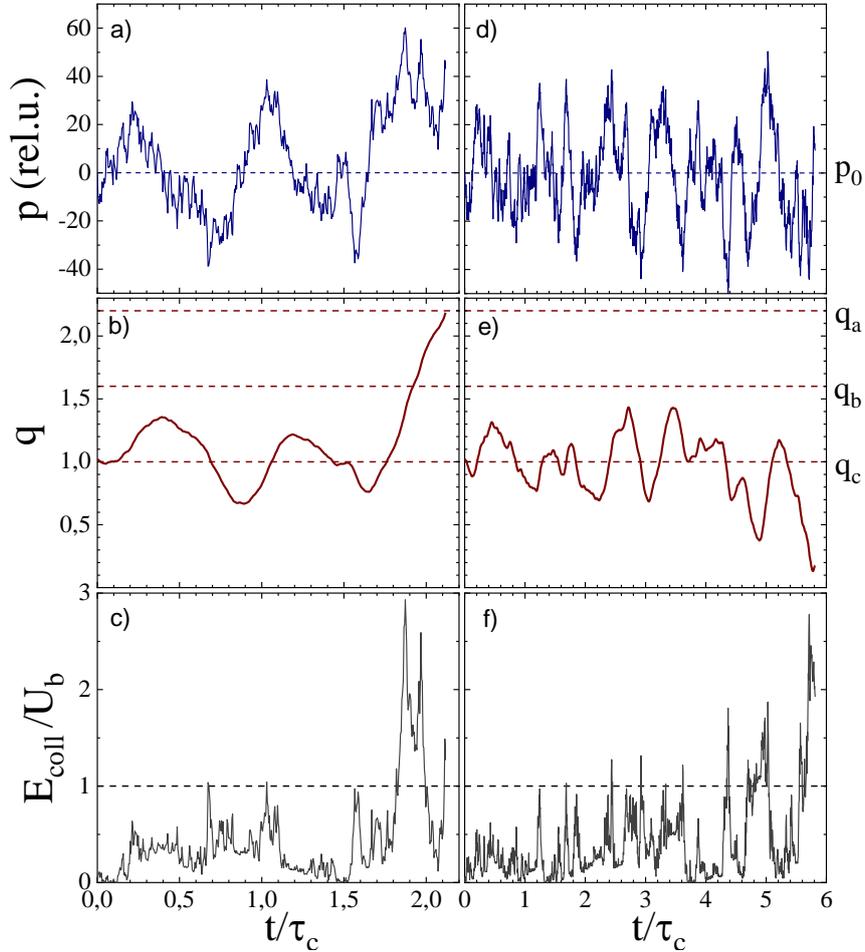

**Figure 2.** Time evolution of the momentum, coordinate, and collective energy of two Brownian particles: the trajectory achieved the absorptive border before the modeling time is expired (panels a-c); the decay of a metastable state does not occur during $t_D$ (panels d-f). $G = 1.5$, $\varphi = 0.721$, $t_D = 5.8\tau_c$, $\tau = 4.15 \cdot 10^{-6}\,\tau_c$.

The right column of Fig. 2 demonstrates the fate of a "survived" trajectory which does not contribute to the decay rate. Here again one sees that the energy of the particle exceeds the barrier height several times, however, the location of the particle in these cases is quite far from the absorbing point and even from the barrier top.

The time-dependent decay rate reads

$$R_{at}(t) = \frac{1}{N_{tot} - N_{at}} \frac{\Delta N_{at}}{\Delta t}. \tag{6}$$

Here $N_{at}$ denotes the number of Brownian particles that reaches $q_a$ by the time moment $t$; $\Delta N_{at}$ is the number of particles reaching $q_a$ during the time lapse $\Delta t$. Typical dependencies $R_{at}(t)$ for different potential shapes and/or values of the damping parameter can be found in many papers (see, e.g., [9,16,23–26]). In Fig. 3 we show the $R_{at}(t)$ dependence obtained in the present work for the cusp potential. This dependence does not seem to be significantly different from that published earlier for other potentials. After some relaxation time, the decay rate attains a quasistationary value $R_D$ which is time independent within the statistical fluctuations. For evaluating $R_D$, we use several bins beginning from the tail of the $R_{at}(t)$-array averaging over them. This algorithm is described in detail in [17,27].

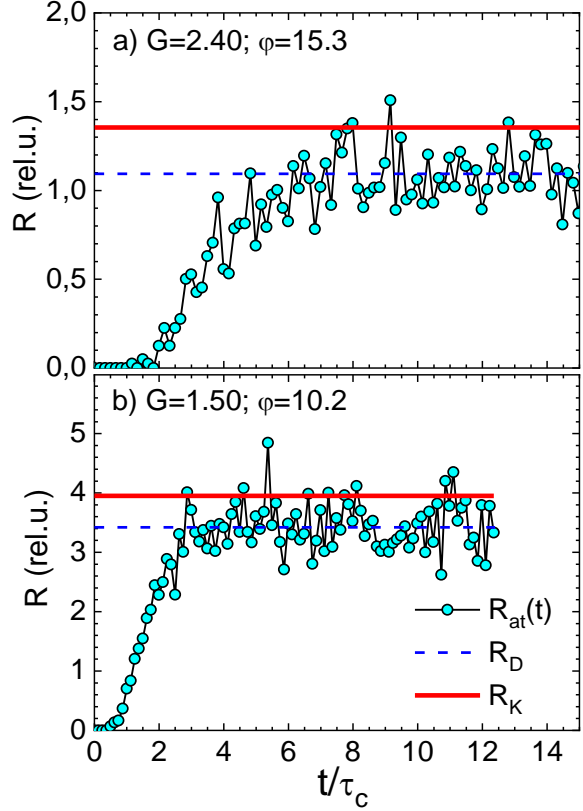

**Figure 3.** The time evolution of the dynamical decay rate (oscillating curve with symbols) with its quasistationary value (thin dashed line) as well as the Kramers rate (thick solid line) for different values of the parameters $G$ and $\varphi$ indicated in the panels.

We choose the number of modeled trajectories and the time lapse of modeling to obtain the quasistationary rate with the relative error not exceeding 2%. The computer code used in the present work was verified by the comparison with the results of [16] for the two-parabolas potential.

This is exactly the quasistationary rate for which the approximate analytical formulas are derived in [11]. In our notations, the Kramers rate for the cusp potential in the overdamping regime reads

$$R_K = \frac{\sqrt{\pi G}}{\varphi \tau_c} \cdot \exp(-G). \qquad (7)$$

As in the case of the two parabolas barrier, this rate (measured in units of $\tau_c^{-1}$) depends only upon the dimensionless scaling parameters $\varphi$ and $G$ [14].

To make a quantitative comparison between $R_D$ and $R_K$, it is convenient to calculate a fractional difference $\xi$:

$$\xi = \frac{R_K - R_D}{R_D}. \qquad (8)$$

## 3. Results

In the present work, we consider a general physical problem, therefore all the results are presented in the dimensionless form. We believe that such a presentation makes our results useful for a wider audience. In Fig. 4 the rates $R_D$ (circles) and $R_K$ (squares) (panel a) and the fractional difference $\xi$ (panel b) are presented as the functions of the damping parameter for two values of $G$, 1.50 (closed symbols) and 3.00 (open symbols). The decrease of the rates with $\varphi$ is in accordance with general expectations since the analytical rate (7) includes the damping parameter in the denominator. The decrease of $\xi$ with $\varphi$ might be easily explained: at smaller values of the damping parameter, the motion is not overdamped anymore. Also, the higher rate for a smaller value of the governing parameter is usual for $G > 1$. A surprising thing in this figure is that the fractional difference does not converge to zero at large values of $\varphi$: the fractional difference seems to reach 10-20% asymptotically. We are not certain about the reason for that, hopefully, it will be investigated in a future publication.

In Fig. 5 we present the dependence of the rates $R_D$ (circles) and $R_K$ (squares) (panel a) and the fractional difference $\xi$ (panel b) upon the governing parameter for two values of $\varphi$. Both values of the damping parameter ($\varphi =10$ and 15) correspond to the overdamping regime, therefore, one expects a good agreement between $R_D$ and $R_K$. However, Fig. 5 shows that this is not the case: as $G$ increases the fractional difference either increases too (for $\varphi = 15$) or behaves in a non-monotonic manner (for $\varphi = 10$). More work is needed to understand this behavior.

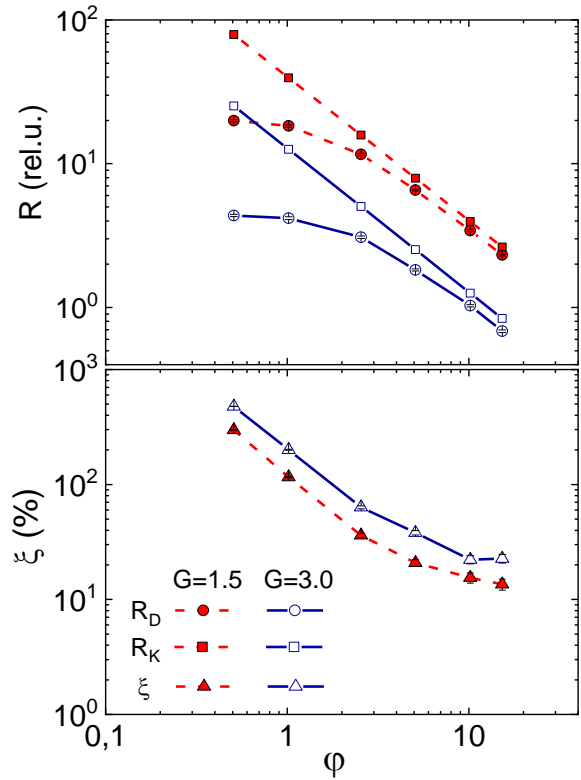

**Figure 4.** The quasistationary (circles) and Kramers (squares) decay rates as the functions of the damping parameter for two values of the governing parameter (panel a). The fractional difference between these two rates (panel b).

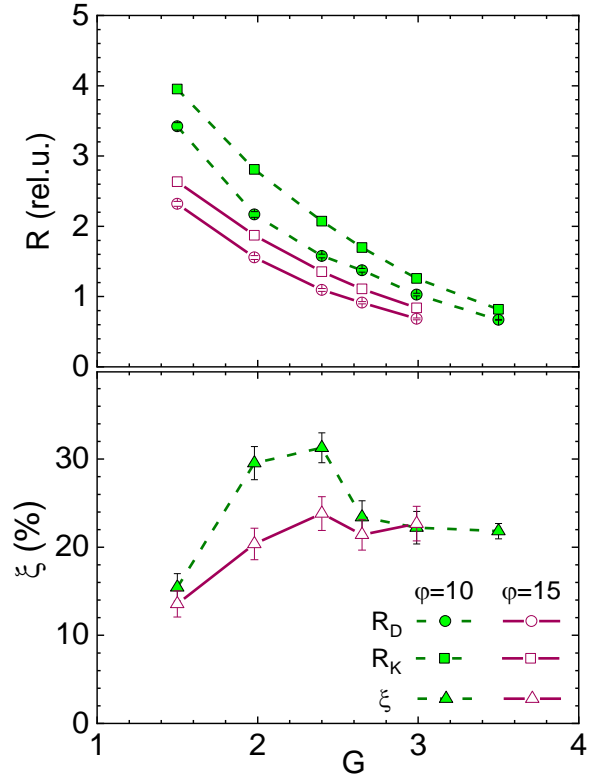

**Figure 5.** The quasistationary (circles) and Kramers (squares) decay rates as the functions of the governing parameter for two values of the damping parameter (panel a). The fractional difference between these two rates (panel b).

## 4. Conclusions

In the present work, for the first time, the accuracy of the Kramers decay rate, $R_K$, for the cusp potential has been explored quantitatively. This is achieved by means of the numerical solution of the stochastic Langevin type equations. These equations are written for the coordinate and its conjugate momentum, i.e. the modeling is performed for the phase space diffusion. The quasistationary decay rate obtained from this modeling is believed to be exact within the framework of the statistical errors which do not exceed 2% in our study.

The Kramers formula for $R_K$ was derived for the overdamping regime, therefore $R_K$ agrees better with the numerical rate $R_D$ at large values of the damping parameter, this is to be expected. An unexpected result is that as the governing parameter $G$ increases (i.e. the barrier becomes higher or the temperature smaller), the agreement between $R_D$ and $R_K$ does not become better. It does become better for the case of the well-studied two parabolas barrier. This contradiction is to be understood in the future. Another option which has not yet been realized is to model the process using the stochastic reduced Langevin equation and/or the corresponding partial differential Smoluchowski equation which are designed for the overdamping regime, i.e. for the spatial diffusion regime. Finally, the integral Kramers formula which seems to be more exact then Eq. (7) can be confronted with the numerical decay rates.

**References**
[1]   Talkner P and Hänggi P 2012 *New Trends in Kramers' Reaction Rate Theory* (Berlin: Springer)
[2]   Weiss G H 1986 Overview of theoretical models for reaction rates *J. Stat. Phys.* **42** 3–36
[3]   Hänggi P, Talkner P and Borkovec M 1990 Reaction-rate theory: fifty years after Kramers

*Rev. Mod. Phys.* **62** 251–341
[4]     Zhou H-X 2010 Rate theories for biologists *Q. Rev. Biophys.* **43** 219–93
[5]     Dudko O K, Hummer G and Szabo A 2006 Intrinsic rates and activation free energies from single-molecule pulling experiments *Phys. Rev. Lett.* **96** 108101
[6]     Chandrasekhar S 1943 Stochastic Problems in Physics and Astronomy *Rev. Mod. Phys.* **15** 1–89
[7]     Jiang Z, Smelyanskiy V N, Isakov S V., Boixo S, Mazzola G, Troyer M and Neven H 2017 Scaling analysis and instantons for thermally assisted tunneling and quantum Monte Carlo simulations *Phys. Rev. A* **95** 012322
[8]     Büttiker M, Harris E P and Landauer R 1983 Thermal activation in extremely underdamped Josephson-junction circuits *Phys. Rev. B* **28** 1268–75
[9]     Nadtochy P N, Kelić A and Schmidt K-H 2007 Fission rate in multi-dimensional Langevin calculations *Phys. Rev. C* **75** 064614
[10]    Ishizuka C, Usang M D, Ivanyuk F A, Maruhn J A, Nishio K and Chiba S 2017 Four-dimensional Langevin approach to low-energy nuclear fission of 236U *Phys. Rev. C* **96** 064616
[11]    Kramers H A 1940 Brownian motion in a field of force and the diffusion model of chemical reactions *Physica* **7** 284–304
[12]    Lavenda B H 1983 Exact solution to Kramers' problem of the escape across a potential barrier in the limit of small resistance *Lett. al Nuovo Cim.* **37** 200–4
[13]    Mel'nikov V I 1991 The Kramers problem: Fifty years of development *Phys. Rep.* **209** 1–71
[14]    Gontchar I I, Chushnyakova M V and Blesman A I 2019 Dimensionless Universal Parameters of the Kramers Problem *J. Phys. Conf. Ser.* **1210** 012052
[15]    Gonchar I I and Kosenko G.I. 1991 Is the Kramers formula applicable for describing the decay of highly excited nuclear systems? *Sov. J. Nucl. Phys.* **53** 133–42
[16]    Gontchar I I, Chushnyakova M V., Aktaev N E, Litnevsky A L and Pavlova E G 2010 Disentangling effects of potential shape in the fission rate of heated nuclei *Phys. Rev. C* **82** 064606
[17]    Gontchar I I and Chushnyakova M V 2017 Thermal decay rate of a metastable state with two degrees of freedom: Dynamical modelling versus approximate analytical formula *Pramana - J. Phys.* **88** 90
[18]    Abkenar M, Gray T H and Zaccone A 2017 Dissociation rates from single-molecule pulling experiments under large thermal fluctuations or large applied force *Phys. Rev. E* **95** 042413
[19]    Bai C, Du L and Mei D 2009 Stochastic resonance induced by a multiplicative periodic signal in a logistic growth model with correlated noises *Open Phys.* **7** 601–6
[20]    Guo F, Zhou Y and Zhang Y 2010 Stochastic Resonance in a Time-Delayed Bistable System Subjected to Dichotomous Noise and White Noise *Chinese J. Phys.* **48** 294–303
[21]    Rosas A, Pinto I L D and Lindenberg K 2016 Kramers' rate for systems with multiplicative noise *Phys. Rev. E* **94** 012101
[22]    Gontchar I I and Chushnyakova M V. 2018 Comment on "Temperature dependence of nuclear fission time in heavy-ion fusion-fission reactions" *Phys. Rev. C* **98** 029801
[23]    Karpov A V., Nadtochy P N, Ryabov E G and Adeev G D 2003 Consistent application of the finite-range liquid-drop model to Langevin fission dynamics of hot rotating nuclei *J. Phys. G Nucl. Part. Phys.* **29** 2365–80
[24]    Pavlova E G, Aktaev N E and Gontchar I I 2012 Modified Kramers formulas for the decay rate in agreement with dynamical modeling *Physica A* **391** 6084–100
[25]    Chushnyakova M V. and Gontchar I I 2018 Thermal decay of a metastable state: Influence of rescattering on the quasistationary dynamical rate *Phys. Rev. E* **97** 032107
[26]    Chushnyakova M V, Gontchar I I and Blesman A I 2018 The Kramers problem in the energy diffusion regime: transient times *J. Phys. Conf. Ser.* **1050** 012018
[27]    Gontchar I I and Krokhin S N 2012 Precision computation of the fission rate of the excited atomic nuclei *Her. Omsk Univ.* **4** 84–7